\documentclass[pra,aps,showpacs,twocolumn,superscriptaddress,amsmath,amssymb]{revtex4}

\usepackage{multirow}
\usepackage{graphicx,dcolumn,bm}
\usepackage{mathrsfs}
\newcommand {\startbild}{\begin{figure}}
\newcommand {\stopbild}{\end{figure}}

\newcommand {\staf}{\begin{equation}}
\newcommand {\stof}{\end{equation}}
\newcommand {\staffeld}{\begin{eqnarray}}
\newcommand {\stoffeld}{\end{eqnarray}}

\newcommand{\ket}[1]{|#1\rangle}
\newcommand{\bra}[1]{\langle #1|}
\renewcommand{\vec}[1]{{\bf #1}}
\begin{document}

\title{Quantum interference and non-locality of independent photons from disparate sources}

\author{R. Wiegner}
\email{ralph.wiegner@physik.uni-erlangen.de}
\affiliation{Institut f\"ur Optik, Information und Photonik, Universit\"at Erlangen-N\"urnberg, Erlangen, Germany}
\homepage{http://www.ioip.mpg.de/jvz/}

\author{J. von Zanthier}
\affiliation{Institut f\"ur Optik, Information und Photonik, Universit\"at Erlangen-N\"urnberg, Erlangen, Germany}

\author{G. S. Agarwal}
\affiliation{Department of Physics, Oklahoma State University, Stillwater, OK, USA}

\date{\today}

\begin{abstract}
We quantitatively investigate the non-classicality and non-locality of a whole new class of mixed disparate quantum and semiquantum photon sources at the quantum-classical boundary. The latter include photon added thermal and photon added coherent sources, experimentally investigated recently by Zavatta \emph{et al.} [\emph{Phys.~Rev.~Lett.~\textbf{103}, 140406 (2009)}]. The key quantity in our investigations is the visibility of the corresponding photon-photon correlation function. We present explicit results on the violations of the Cauchy-Schwarz inequality - which is a measure of nonclassicality - as well as of Bell-type inequalities.

\end{abstract}

\maketitle

\section{Introduction}

The question of interference between independent photons has attracted considerable attention since the celebrated work of Hong, Ou and Mandel who showed how two photons of the pair emitted in a non-linear crystal by the process of spontaneous parametric down conversion (SPDC) can interfere on a beam splitter~\cite{HOM}. Many subsequent investigators verified the Hong-Ou-Mandel interference using a variety of SPDC sources \cite{HOM1,HOM2,HOM3}. Also the interference of photons from truly independent quantum sources, e.g., two distant atoms, has been analyzed in great detail \cite{mandel,GSA02} and recently observed~\cite{Grangier06,Zeilinger06,Monroe07,Eschner09}. Note that the interference is displayed in terms of the photon-photon correlation function \cite{glauber1} rather than in measurements of the mean intensity. So far, most works have concentrated on the interference produced by independent but identical sources. However, the questions then arise: what is the extent to which photons from independent but disparate sources can interfere \cite{shields}? Further, what is the nature of the radiation field generated and to what extent are the properties of the field strictly quantum? Can one use such a quantum character to shed light on the non-local character of the field? We provide quantitative answers to these questions. Specifically we investigate the quantum interference of two disparate sources, one being a single photon source like an excited atom and the other a source which can give rise to properties ranging from completely classical to completely quantum. In the latter category we specifically examine  sources which are either in a state called photon added coherent state \cite{PACS} or photon added thermal state \cite{PATS}. The quantum properties of these sources arise from the fact that one has added photons to classical states like thermal and coherent states. Such sources have been recently experimentally realized and their properties studied \cite{abstract2,PACSexp,PACSexp1,PATSexp}. We calculate the interference pattern of the corresponding photon-photon correlation functions, derive their visibilities and discuss the conditions under which the photon-photon correlation functions displays strictly quantum behavior. The latter is formulated quantitatively in terms of the Cauchy-Schwarz (CS) inequality \cite{Loudon80} valid for classical fields. After revealing the quantum nature of the light fields generated by disparate sources we discuss the conditions under which Bell inequalities can be violated \cite{bell,bell1,chsh,EPRreview}.

The organization of the paper is as follows: In Sect.~\ref{sG2mixed} we derive the photon-photon correlation functions and the corresponding visibilities for combinations of mixed disparate quantum-classical and quantum-semiquantum light sources. As quantum source we consider throughout the paper a single photon source, e.g., an initially excited two-level atom spontaneously emitting a single photon. As second source we assume either a purely classical source exhibiting Poissonian or thermal statistics \cite{QMxC1,QMxC2,QMxT} or a semiquantum source emitting a field which is either in a photon added coherent state (PACS) or a photon added thermal state (PATS). In Sect.~\ref{sschwarz} we formulate a version of the CS inequality for position dependent photon-photon correlation functions in order to investigate quantitatively when the radiation fields display non-classicality, by explicitly showing the regimes where the CS inequality is violated. In Sect.~\ref{schsh} we then discuss, under which conditions Bell inequalities can be violated by our mixed quantum-classical and quantum-semiquantum sources. In Sect.~\ref{summary} we conclude with a summary of our results.

\section{Photon-Photon correlations and the visibility of interference}
\label{sG2mixed}

\begin{figure}[h!]
\centering
\includegraphics[width=0.4 \textwidth]{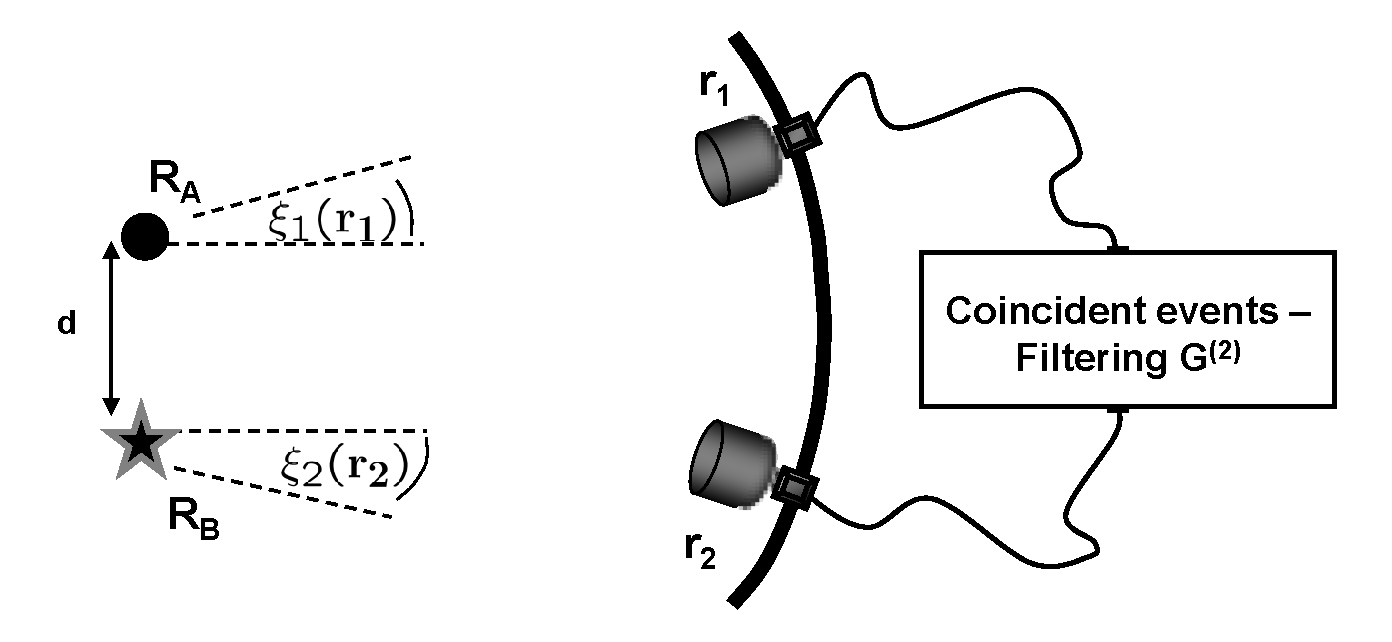}
\caption{\label{classquant} Setup of the investigated arrangement. We consider one photon source localized at position ${\bf R}_A$ and another photon source at position ${\bf R}_B$. Using two detectors at positions ${\bf r}_1$ and ${\bf r}_2$ we assume that each detector registers a photon in the far field of the sources.}
\end{figure}

To derive the photon-photon correlation function of the fields emitted by the various mixed disparate quantum-classical and quantum-semiquantum sources, we consider two sources $A$ and $B$ localized at positions ${\bf R}_A$ and ${\bf R}_B$, respectively, with a separation $d >\!> \lambda$, so that any interaction between them can be neglected (cf.~Fig. \ref{classquant}). We define a measurement cycle by two detection events at two detectors, i.e., we assume that each of the two detectors registers one photon so that two photon absorption processes at one detector can be excluded. The detectors are located at positions ${\bf r}_1$ and ${\bf r}_2$, where $|\vec{r}_i - \vec{R}_n| \gg d$ ($i = 1,2$, $n = A, B$), so that the far field condition is fulfilled. To simplify further calculations we only consider coincident detections, i.e., simultaneous events at both detectors.

In Glauber notation \cite{glauber1} the photon-photon correlation function can be written as 

\begin{eqnarray}\label{tutg1}
&&G^{(2)}({\bf r}_1,{\bf r}_2)=\nonumber\\
&&\langle \hat{{\bf E}}^{(-)}_{A+B}\!({\bf r}_1)\hat{{\bf E}}^{(-)}_{A+B}\!({\bf r}_2)\hat{{\bf E}}^{(+)}_{A+B}\!({\bf r}_2)\hat{{\bf E}}^{(+)}_{A+B}\!({\bf r}_1)\rangle,
\end{eqnarray}

\noindent where $(\hat{{\bf E}}^{(-)}_{A+B})^\dagger = \hat{{\bf E}}^{(+)}_{A+B}$ and $\langle\ldots \rangle$ denotes the expectation value corresponding to a temporal average for stationary fields. Hereby, the positive frequency part of the electric field operator $\hat{{\bf E}}^{(+)}_{A+B}({\bf r}_i)$ takes the form \cite{Agarwal74}

\begin{eqnarray}
\label{tutg2}
\hat{{\bf E}}^{(+)}_{A+B}({\bf r}_i) \equiv \hat{{\bf A}}^{(+)}_i +  \hat{{\bf B}}^{(+)}_{i}\propto \hspace{1cm}\nonumber\\
\hspace{0.5cm} e^{-ik({\bf \hat{r}}_i\,{\bf R}_A)}\,{s} + e^{-ik({\bf \hat{r}}_i\,{\bf R}_B)}\,{a},
\end{eqnarray}
where $k=\frac{2\pi}{\lambda}=\frac{\omega}{c}$ denotes the wave number of the two sources, assumed to radiate at the same frequency $\omega$ and ${\bf \hat{r}}_i:=\frac{{\bf r}_i}{|{\bf r}_i|}$ is a unit vector in the direction of the $i$th detector. 

Except for the next paragraph, we consider in this paper for source $A$ a single photon source, e.g., a two-level atom with upper level $\ket{e}$ and ground state $\ket{g}$. The atom is assumed to be initially fully excited to the state $\ket{e}$ so that after the spontaneous emission of a single photon the atom is transferred to the state $\ket{g}$. In this case the operator ${s}$ in Eq. (\ref{tutg2}) denotes the lowering operator $\ket{g}\bra{e}$. As source $B$ we consider in the following either a more traditional source like a coherent source or a thermal source \cite{QMxC1,QMxT,QMxC2} or a source which displays both, classical and quantum properties. For the latter we  specifically investigate the configurations where source $B$ is either in a PACS \cite{PACS} or in a PATS \cite{PATS}. The operator ${a}$ in Eq. (\ref{tutg2}) thus denotes the photon annihilation operator for the source $B$. 

For a better comparison with the following calculations we start to review the pure classical case where the two independent light sources $A$ and $B$ emit coherent light \cite{mandel}. The operator $s$ in Eq.~(\ref{tutg2}) denotes in this case the photon annihilation operator for the source $A$. Since the two light fields are in a coherent state, $|\alpha\rangle_A$ and $|\alpha\rangle_B$ with mean photon numbers $\bar{n}_A$ and $\bar{n}_B$, respectively, we calculate from Eqs.~(\ref{tutg1}) and (\ref{tutg2}) the photon-photon correlation function $G^{(2)}_{{ Class}}(\vec{r}_1, \vec{r}_2)$ to 
\staffeld
\label{G2calcCC}
G^{(2)}_{{ Class}}(\vec{r}_1, \vec{r}_2) = \bar{n}_A^2 + 2 \, \bar{n}_A\,\bar{n}_B + \bar{n}_B^2 +\hspace{1cm}\nonumber\\
\hspace{2cm} + 2 \, \bar{n}_A\,\bar{n}_B\,\cos{(\varphi_2 - \varphi_1)}\,,
\stoffeld
where the relative phase $\varphi_i$ is given by $\varphi_i \equiv \varphi({\bf r}_i)= k\,d\,\sin(\xi(\vec{r}_i))$ (cf.~Fig.~\ref{classquant}). The visibility
\staf
{\cal V} := \frac{G^{(2)}_{max} - G^{(2)}_{min}}{G^{(2)}_{max} + G^{(2)}_{min}}
\stof
of this classical signal calculates to
\staf
\label{VCC}
{\cal V}_{{ Class}} = \frac{2 \, \bar{n}_A\,\bar{n}_B}{(\bar{n}_A + \bar{n}_B)^2}.
\stof
If one beam has a much higher mean photon number than the other, e.g., $\bar{n}_A \gg \bar{n}_B$, the visibility ${\cal V}_{{ Class}}$ goes to zero as $\frac{2\,\bar{n}_B}{\bar{n}_A}$. On the other hand, for $\bar{n}_A = \bar{n}_B$, we obtain the maximal value ${\cal V}_{{ Class}} = 50\%$ \cite{mandel}.

Using Eq.~(\ref{tutg1}) for mixed quantum-classical or mixed quantum-semiquantum sources the photon-photon correlation function is found to be


\begin{widetext}
\begin{eqnarray}
\label{G2calc}
G^{(2)}(\vec{r}_1, \vec{r}_2) = \langle (\hat{{\bf A}}^{(-)}_1 +  \hat{{\bf B}}^{(-)}_{1})(\hat{{\bf A}}^{(-)}_2 +  \hat{{\bf B}}^{(-)}_{2}) \, (\hat{{\bf A}}^{(+)}_2 +  \hat{{\bf B}}^{(+)}_{2})(\hat{{\bf A}}^{(+)}_1 +  \hat{{\bf B}}^{(+)}_{1})\rangle = \langle \hat{{\bf A}}^{(-)}_1 \hat{{\bf A}}^{(+)}_1 \rangle \langle \hat{{\bf B}}^{(-)}_2 \hat{{\bf B}}^{(+)}_2 \rangle +\nonumber \\
+ \langle \hat{{\bf B}}^{(-)}_1 \hat{{\bf B}}^{(+)}_1 \rangle \langle \hat{{\bf A}}^{(-)}_2 \hat{{\bf A}}^{(+)}_2 \rangle + \langle \hat{{\bf A}}^{(-)}_1 \hat{{\bf A}}^{(+)}_2 \rangle \langle \hat{{\bf B}}^{(-)}_2 \hat{{\bf B}}^{(+)}_1 \rangle + \langle \hat{{\bf B}}^{(-)}_1 \hat{{\bf B}}^{(+)}_2 \rangle\langle \hat{{\bf A}}^{(-)}_2 \hat{{\bf A}}^{(+)}_1 \rangle + \langle \hat{{\bf B}}^{(-)}_1 \hat{{\bf B}}^{(-)}_2 \hat{{\bf B}}^{(+)}_2 \hat{{\bf B}}^{(+)}_1 \rangle
\end{eqnarray}
\end{widetext}

where we have used the fact that the two photon sources are uncorrelated. In difference to the pure classical case above we can omit in Eq.~(\ref{G2calc}) the term $\langle \hat{{\bf A}}^{(-)}_1 \hat{{\bf A}}^{(-)}_2 \hat{{\bf A}}^{(+)}_2 \hat{{\bf A}}^{(+)}_1 \rangle$ as it vanishes identically for single photon emitters. However, unlike the pure quantum mechanical signal~\cite{ou} consisting of, e.g., two single photons scattered by two atoms, we additionally have to take into account in Eq.~(\ref{G2calc}) the term $\langle \hat{{\bf B}}^{(-)}_1 \hat{{\bf B}}^{(-)}_2 \hat{{\bf B}}^{(+)}_2 \hat{{\bf B}}^{(+)}_1 \rangle$. This term represents the probability that source $B$ has emitted two photons which are subsequently measured at the two detectors. In the following, we start to discuss the configuration where source $B$ is a standard classical source (Sect.~\ref{class1}). Thereafter, we study the configuration where source $B$ is either in a PACS or PATS, i.e., in a state at the quantum-classical border (Sect.~(\ref{semiquantum1})). These states are particularly interesting as they display properties ranging from completely classical to completely quantum.

\subsection{Source $B$ as a classical photon source}
\label{class1}
\subsubsection{coherent source}
\label{secQMxC}

Let us start to consider source $B$ to be in a coherent state $\ket{\alpha}$. Expanded in terms of Fock states $\ket{n}$, these states take the well known form \cite{glauber1}

\staf
\label{coherentstate}
\ket{\alpha} = e^{- \left|\alpha \right|^2 /2} \sum_{n = 0}^{\infty} \frac{\alpha^n}{\sqrt{n !}} \ket{n},
\stof 

\noindent where $\alpha$ is any complex number. The $k$-th moment of the photon number operator $a^{\dagger}\,a$ in the coherent beam can be calculated to 

\staf
\label{momc}
\left\langle {a}^{\dagger k}\,{a}^k \right\rangle_C = \bra{\alpha} {a}^{\dagger k}\,{a}^k \ket{\alpha} = \left| \alpha \right|^{2\,k} = \bar{n}^k \,,
\stof

\noindent where $\bar{n} = \left\langle a^{\dagger}\,a \right\rangle \propto I_B \propto |E_B|^2$ is the mean photon number with $E_B$, the amplitude of the electric field of source $B$. Using Eq.~(\ref{G2calc}), the photon-photon correlation function of one quantum and one coherent source is then found to be

\begin{eqnarray}
\label{G2calcC}
G^{(2)}_{{ C}}(\varphi_1, \varphi_2) = \bar{n}^2 + 2\,\bar{n}\, (1 + \cos{(\varphi_2 - \varphi_1)})\,,
\end{eqnarray}

\noindent so that the visibility of the mixed quantum-coherent photon-photon correlation function $G^{(2)}_{{ C}}(\varphi_1, \varphi_2)$ becomes

\staf
\label{VQMC}
{\cal V}_{{ C}} = \frac{1}{1 + \frac{\bar{n}}{2}}\,.
\stof

\noindent For a classical coherent plane wave with $\bar{n} \ll 1$  we obtain ${\cal V}_{{ C}} = 100\%$, for $\bar{n} = 1$ we arrive at ${\cal V}_{{ C}} = 66\%$, and for $\bar{n} \rightarrow \infty$ we have ${\cal V}_{{ C}} = 0$. In \cite{QMxC1} a single photon created by SPDC was entangled with an attenuated laser beam on a beam splitter. The authors obtained a modulation depth of $(84 \pm 3.2)\%$. According to Eq.~(\ref{VQMC}) this corresponds to a mean photon number in the coherent beam of $0.293 < \bar{n} < 0.475$.

\subsubsection{thermal source}
\label{secQMxT}

Now, let us assume that source $B$ exhibits thermal statistics. The corresponding density operator expanded in Fock-states $\ket{n}$ takes the form

\staf
\label{rhot}
\hat{{\bf \rho}}_T = \sum_{n = 0}^\infty \frac{\bar{n}^n}{(\bar{n} + 1)^{n + 1}} \ket{n}\bra{n}\,.
\stof

\noindent For the moments $\left\langle {a}^{\dagger k}\,{a}^k \right\rangle_T$ one calculates

\staf
\label{momt}
\left\langle {a}^{\dagger k}\,{a}^k \right\rangle_T = Tr\left[ {a}^{\dagger k}\,{a}^k \,\hat{{\bf \rho}}_T \right] =  k!\,\bar{n}^k\,,
\stof

\noindent With these expressions Eq.~(\ref{G2calc}) takes the form

\begin{eqnarray}
\label{G2calcT}
G^{(2)}_{{ T}}(\varphi_1, \varphi_2) = 2\,\bar{n}^2 + 2\,\bar{n}\, (1 + \cos{(\varphi_2 - \varphi_1)})\,.
\end{eqnarray}

\noindent The visibility of the mixed quantum-thermal signal $G^{(2)}_{{ T}}(\varphi_1, \varphi_2)$ thus calculates to

\staf
\label{VQMT}
{\cal V}_{{ T}} = \frac{1}{1 + \bar{n}}\,,
\stof

\noindent what leads again to ${\cal V}_{{ T}} = 100\%$ for $\bar{n} \ll 1$ and to ${\cal V}_{{ T}} = 0$ for $\bar{n} \rightarrow \infty$. If we choose $\bar{n} = 1$ we obtain ${\cal V}_{{ T}} = 50\%$, corresponding to the visibility of a pure classical photon-photon correlation signal exhibited by coherent light fields if $\bar{n}_A = \bar{n}_B$ (cf.~Eq.~(\ref{VCC})). In a recent experiment a visibility of $82\%$ was measured in this configuration, equivalent to a mean photon number in the thermal beam of $\bar{n} = 0.22$~\cite{QMxT}.

\subsection{Source $B$ as a semiquantum photon source}
\label{semiquantum1}
\subsubsection{PACS}
\label{secQM}

Now, let us consider the case where the field of source $B$ is in a photon added coherent state~\cite{PACS}. In terms of the coherent states $\ket{\alpha}$ a normalized \textit{single} photon added coherent state (in the following abbreviated $PC$) can be written as

\staf
\label{PACSstate}
\ket{\alpha, 1} = \frac{{a}^{\dagger}\,\ket{\alpha}}{\sqrt{1 + \bar{n}}}\;.
\stof

\noindent Hereby, $\bar{n}$ corresponds to the mean photon number in the coherent part of the light field. The k-th moments of the photon number operator $a^{\dagger}\,a$ then take the form

\staf
\label{momspacs}
\bra{\alpha, 1} {a}^{\dagger k}\,{a}^k \ket{\alpha, 1} = \frac{\bra{\alpha} {a}\,{a}^{\dagger k}\,{a}^{k}\,{a}^{\dagger}\, \ket{\alpha}}{1 + \bar{n}} \, ,
\stof

\noindent so that the first and the second moments read  

\staffeld
\label{momspacs1}
\mathfrak{n}_{PC} \equiv  \left\langle {a}^{\dagger}\,{a} \right\rangle_{PC}  &=& \frac{\bar{n}^2 + 3\,\bar{n} + 1}{1 + \bar{n}}\,, \nonumber \\
\left\langle {a}^{\dagger 2}\,{a}^2 \right\rangle_{PC} &=& \bar{n}^2 + 4\,\bar{n}\,.
\stoffeld
where we have introduced the net photon number $\mathfrak{n}_{PC}$ of the PACS field. With the help of Eq.~(\ref{G2calc}), we arrive at

\begin{eqnarray}
\label{G2calcPACS}
G^{(2)}_{{ PC}}(\varphi_1, \varphi_2) =  \bar{n}^2 + 4\,\bar{n} +\hspace{2.5cm}\nonumber\\
+\, 2\,\frac{\bar{n}^2 + 3\,\bar{n} + 1}{1 + \bar{n}}\,(1 + \cos{(\varphi_2 - \varphi_1)})\,,
\end{eqnarray}

\noindent and the visibility ${\cal V}_{{ PC}}$ thus takes the form

\staf
\label{VQMPACS}
{\cal V}_{{ PC}} = \frac{1}{1 + \frac{\bar{n}^3 + 5\,\bar{n}^2 + 4\,\bar{n}}{2\,(\bar{n}^2 + 3\,\bar{n} + 1)}}\,.
\stof

In case of a mixed quantum-coherent or mixed quantum-thermal photon-photon correlation signal (cf.~Eqs.~(\ref{G2calcC}) and (\ref{G2calcT})) the mean photon number $\bar{n}$ in the coherent or thermal beam of source $B$ trivially corresponds to the net photon number of the field, $\mathfrak{n}_{C} =  \bar{n}$ and $\mathfrak{n}_{T} =  \bar{n}$, respectively. However, if we want to compare the visibility ${\cal V}_{{ PC}}$ of the mixed quantum-PACS photon-photon correlation signal to the foregoing results (cf.~Eqs.~(\ref{VQMC}) and (\ref{VQMT})), we have to express ${\cal V}_{{ PC}}$ in terms of the net photon number $\mathfrak{n}_{PC}$. We can rewrite Eq.~(\ref{VQMPACS}) by using the identity (cf.~Eq.~(\ref{momspacs1}))

\staffeld
\label{nPACS}
\bar{n} = -\frac{3}{2} + \frac{1}{2} \mathfrak{n}_{PC} + \hspace{2.5cm} \nonumber\\
\hspace{1cm}+ \frac{1}{2} \sqrt{5 - 2\mathfrak{n}_{PC} + \mathfrak{n}_{PC}^2}
\stoffeld

\noindent and arrive at a visibility ${\cal V}_{{PC}}$ depending on $\mathfrak{n}_{PC}$ rather than $\bar{n}$. Since the analytic expression is rather complex, we do not present the explicit result - however, the corresponding outcome is plotted in Fig.~\ref{visQMC}.  

For $\mathfrak{n}_{PC} \rightarrow 1$ we obtain ${\cal V}_{{PC}} = 100\%$, whereas for $\mathfrak{n}_{PC} \rightarrow \infty$ we have ${\cal V}_{{PC}} = 0$, like in the mixed quantum-classical cases above. 
However, the visibility ${\cal V}_{{PC}}$ of the mixed quantum-PACS photon-photon correlation function always displays higher values for a given net photon number $\left\langle a^{\dagger}\,a \right\rangle$ than the visibility ${\cal V}_{{C}}$ for a mixed quantum-coherent source. For example, for a net photon number $\mathfrak{n}_{PC} = 1$, we obtain a visibility ${\cal V}_{{ PC}} = 100\%$ in contrast to a visibility ${\cal V}_{{ C}} =  66\%$ for a net photon number $\mathfrak{n}_{C} = 1$ in case of a mixed quantum-coherent source.

\subsubsection{PATS}

Next, we consider the field of source $B$ to be in a PATS \cite{PATS}. In case of a \textit{single} photon added thermal state (in the following abbreviated as ST), the normalized density operator $\hat{{\bf \rho}}_{PT}$ can be written in the Fock basis as

\staffeld
\label{rhopt}
\hat{{\bf \rho}}_{PT} &=& \frac{1}{\bar{n} + 1}\,{a}^\dagger\, \hat{{\bf \rho}}_T\,{a} = \nonumber\\
&=& \frac{1}{\bar{n}\,(\bar{n}+1)}\,\sum_{n = 0}^\infty \left(\frac{\bar{n}}{\bar{n} + 1}\right)^n n\,\ket{n}\bra{n}\,,
\stoffeld

\noindent where $\bar{n}$ corresponds to the mean photon number of the thermal part of the light field. In contrast to the thermal density operator of Eq.~(\ref{rhot}) it is obvious that the vacuum term is missing and higher excited terms are rescaled. For the first and second moments of the photon number operator $a^{\dagger}\,a$ we obtain 



\staffeld
\label{momspats}
\mathfrak{n}_{PT} \equiv   \left\langle {a}^{\dagger}\,{a} \right\rangle_{PT}  &=& 2\,\bar{n} + 1\,, \nonumber \\
\left\langle {a}^{\dagger 2}\,{a}^2 \right\rangle_{PT} &=& 6\,\bar{n}^2 + 4\,\bar{n}
\stoffeld
where we have introduced again the net photon number $\mathfrak{n}_{PT}$ of the PATS field of source $B$. Thus Eq.~(\ref{G2calc}) calculates to

\begin{eqnarray}
\label{G2calcPATS}
G^{(2)}_{{ PT}}(\varphi_1, \varphi_2) = \frac{6\,\bar{n}^3 + 10\,\bar{n}^2 + 4\,\bar{n}}{\bar{n} + 1} + \hspace{2cm}\\ \nonumber
\hspace{1cm}+\, 2\,(2\,\bar{n} + 1)(1 + \cos{(\varphi_2 - \varphi_1)}) \,,
\end{eqnarray}

\noindent and for the visibility ${\cal V}_{{ PT}}$ we derive

\staf
\label{VQMPATS}
{\cal V}_{{ PT}} = \frac{1}{1 + \frac{6\,\bar{n}^3 + 10\,\bar{n}^2 + 4\,\bar{n}}{2\,(\bar{n} + 1)(2\,\bar{n} + 1)}}\,.
\stof

\noindent Again, we have to formulate the visibility ${\cal V}_{{ PT}}$ in terms of the net photon number $\mathfrak{n}_{PT}$ in order to compare it to the previous mixed quantum-classical results. Inverting~Eq.~(\ref{momspats})

\staf
\label{nPATS}
\bar{n} = \frac{\mathfrak{n}_{PT} - 1}{2}
\stof

\noindent we obtain 

\staf
\label{VQMPATSnet}
{\cal V}_{{ PT}} = \frac{4\,\mathfrak{n}_{PT}}{3\,\mathfrak{n}_{PT}^2+2\,\mathfrak{n}_{PT}-1}\,.
\stof

\noindent From Eq.~(\ref{VQMPATSnet}) we can see that for $\mathfrak{n}_{PT} \rightarrow 1$ the visibility ${\cal V}_{{ PT}}$ goes to $100\%$ whereas for $\mathfrak{n}_{PT} \rightarrow \infty$ we have ${\cal V}_{{ PT}} = 0$. Moreover, again the visibility ${\cal V}_{{ PT}}$ is taking higher values for any net photon number than ${\cal V}_{{ T}}$ of the mixed quantum-thermal state (cf.~Eq.~(\ref{VQMT})). If we choose for example $\mathfrak{n}_{PT} = 2$ we obtain ${\cal V}_{{ PT}} = 53.3\%$, whereas  for $\mathfrak{n}_{T} = 2$ we get ${\cal V}_{{ T}} = \frac{1}{3}$  (cf.~Eqs.~(\ref{momt}) and (\ref{VQMT})). 

\begin{figure}
\centering
\includegraphics[width=0.4 \textwidth]{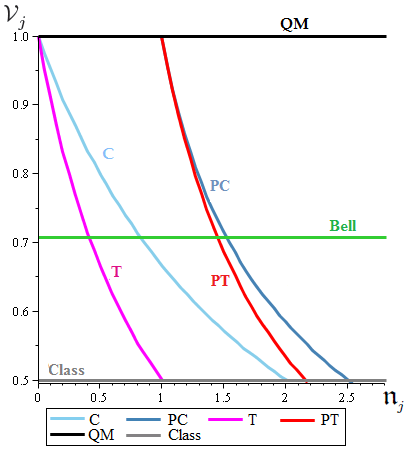}
\caption{\label{visQMC} Plot of the visibility of the photon-photon correlation function for different light sources dependent on the net photon number $\mathfrak{n}_{j}$ ($j = \{T,PT,C,PC\}$). \emph{QM} (\emph{Class}) abbreviates the visibility of a pure quantum (classical) source. \emph{Bell} depicts the border to violate locality (cf.~section \ref{schsh} for details).}
\end{figure}

Fig.~\ref{visQMC} displays the visibility of the various photon-photon correlation functions as a function of the net photon numbers $ \mathfrak{n}_{j}$ ($j = \{T,PT,C,PC\}$) for the investigated combinations of mixed quantum-classical and mixed quantum-semiquantum sources. It thus shows a summary of the results discussed in this section. For comparison, the constant visibility ${\cal V} = 1$ obtained for a pure quantum signal, e.g., two photons from two single photon emitters, and the constant visibility obtained for a pure classical signal (${\cal V}_{{Class}} = \frac{1}{2}$) are also plotted.

\section{Test of the Quantum Character of the mixed radiation fields}
\label{sschwarz}

After deriving the various photon-photon correlation signals of the mixed quantum-classical and quantum-semiquantum sources in section \ref{sG2mixed}, the natural question arises of how to characterize and specify the nonclassicality \cite{vogel1,Luis09} of these sources. For this purpose we introduce a particular version of the CS inequality for our photon-photon correlations, valid for classical fields. If this inequality is violated the underlying radiation field is non-classical. This is because the key ingredient in the derivation of the CS inequality is the assumption that the Glauber-Sudarshan P-function \cite{glauber1,sudarshan} behaves like a classical probability distribution. In its simplest form it reads~\cite{Loudon80}

\staf
\label{27}
\left\langle a^{\dagger \,2}\,a^2\right\rangle \,\left\langle b^{\dagger\,2}\,b^2\right\rangle \geq \left|\left\langle a^{\dagger}\,b^{\dagger}\,a\,b \right\rangle\right|^2\,,
\stof

\noindent where $a$ and $b$ are the mode variables for the classical fields. For quantum fields, $a$ and $b$ are the annihilation operators. In this case the inequality can be violated since the Glauber-Sudarshan P-function can be negative, singular or need not exist. 
However, for the two point photon-photon correlations of our sources we have to derive an alternate form of the CS inequality. As discussed in Appendix \ref{csapp} a direct application of Eq.~(\ref{27}) does not work.

Let us consider the net intensity $I(\varphi)$ of two sources $A$ and $B$ at the point $\varphi(\vec{r})$. If the underlying probability distribution (represented by the P-function) is positive, then the expression

\staf
\label{trrhodeltai}
\left\langle : [\alpha ( \hat{I}(\varphi_1) - \langle \hat{I}(\varphi_1)\rangle) + \beta ( \hat{I}(\varphi_2) - \langle \hat{I}(\varphi_2)\rangle) ]^2 :\right\rangle
\stof
\noindent should be positive for arbitrary $\alpha$ and $\beta$ (where $: \;:$ abbreviates normal ordering). As shown in the appendix the corresponding photon-photon correlations must satisfy this inequality provided that the P-function is positive. Any violation of this inequality implies then that the underlying radiation field is a nonclassical field. From Eq.~(\ref{trrhodeltai}) we arrive at the inequality (see Appendix \ref{csapp})


\staf
\label{schwarzeq}
{\cal S} := \frac{\left|\tilde{G}^{(2)}(\varphi_1, \varphi_2)\right|^2}{\tilde{G}^{(2)}(\varphi_1, \varphi_1)\,\tilde{G}^{(2)}(\varphi_2, \varphi_2)} \leq 1\,,
\stof

\noindent where the variance $\tilde{G}^{(2)}(\varphi_k, \varphi_l)$ is given by

\staf
\tilde{G}^{(2)}(\varphi_k, \varphi_l) := G^{(2)}(\varphi_k, \varphi_l) - \left\langle I_A + I_B \right\rangle^2\,,
\stof

\noindent and $G^{(2)}_j(\varphi_k, \varphi_l)$ ($j = \{T,PT,C,PC\}$ corresponds to the photon-photon correlation functions given in Eqs.~(\ref{G2calcCC}), (\ref{G2calcC}), (\ref{G2calcT}), (\ref{G2calcPACS}) and (\ref{G2calcPATS}). Hereby, $I_A$ ($I_B$) abbreviates the intensity of the photon sources located at ${\bf R}_A$ (${\bf R}_B$) (cf.~Fig. \ref{classquant}). Keeping in mind that we suppose uncorrelated sources and that we have $\left\langle I_A \right\rangle_{QM} = 1$, and taking into account the corresponding moments given in Eqs.~(\ref{momc}), (\ref{momt}), (\ref{momspacs}) and (\ref{momspats}), the variances $\tilde{G}^{(2)}_{{ j}}(\varphi_k, \varphi_l)$ are calculated to

\staffeld
\label{scc}
&&\tilde{G}^{(2)}_{{ Class}}(\varphi_k, \varphi_l) = 2\,\bar{n}^2\,c_{kl}\,,\\
\label{sc}
&&\tilde{G}^{(2)}_{{ C}}(\varphi_k, \varphi_l) = 2\,\bar{n}\,c_{kl} - 1\,,\\
\label{st}
&&\tilde{G}^{(2)}_{{ T}}(\varphi_k, \varphi_l) = \bar{n}^2 + 2\,\bar{n}\,c_{kl} - 1\,,\\
\label{sqc}
&&\tilde{G}^{(2)}_{{PC}}(\varphi_k, \varphi_l) = \frac{-(3\,\bar{n}^2 + 4\,\bar{n} + 2)}{(\bar{n} + 1)^2} +\hspace{1cm}\nonumber\\
&&\hspace{1cm} +\,\frac{(2\bar{n}^3+8\bar{n}^2+8\bar{n}+2)c_{kl}}{(\bar{n} + 1)^2} \,,\\
\label{sqp}
&&\tilde{G}^{(2)}_{{PT}}(\varphi_k, \varphi_l) = 2\,\bar{n}^2 + 2\,(2\bar{n}+1)c_{kl} - 2\,,\\
\label{sqq}
&&\tilde{G}^{(2)}(\varphi_k, \varphi_l) = 2\,c_{kl} -2\,,
\stoffeld


\noindent with the abbreviation \mbox{$c_{kl} \equiv \cos{(\varphi_{k} - \varphi_{l})}$}. Note that for comparison we have also included the pure quantum mechanical case, i.e., two single photon emitters, where 

\staf
\label{gqq}
G^{(2)}(\varphi_1, \varphi_2) = 2\, (1 + \cos{(\varphi_2 - \varphi_1)})
\stof

\noindent with ${\cal V} = 1$~\cite{GSA02}. We emphasize again that in case of a mixed quantum-PACS field or a mixed quantum-PATS field, we have to write $\tilde{G}^{(2)}_{{ PC}}$ and $\tilde{G}^{(2)}_{{ PT}}$ as a function of the net photon numbers $\mathfrak{n}_{PC}$ and $\mathfrak{n}_{PT}$, respectively, to appropriately compare these signals to the mixed quantum-classical photon-photon correlation signals. To this end we have to insert the identities for the net photon numbers of Eq.~(\ref{nPACS}) and (\ref{nPATS}) in Eqs.~(\ref{sqc}) and (\ref{sqp}), respectively. If we now plug the variances of Eqs.~(\ref{scc})-(\ref{sqq}) in the CS inequality Eq.~(\ref{schwarzeq}), we obtain a violation for every mixed quantum-classical and mixed quantum-semiquantum signal below a certain net photon number $\left\langle a^{\dagger}\,a \right\rangle_j$ ($j = \{T,PT,C,PC\}$), after an optimization with respect to $\varphi_{1}$ and $\varphi_{2}$, i.e., with respect to the detector positions. Note that according to Fig.~\ref{visQMC}, reducing the net photon number $\mathfrak{n}_{j}$ in the field of source $B$ increases the visibility ${\cal V}_{{j}}$ of the corresponding photon-photon correlation signal ($j = \{T,PT,C,PC\}$).

\begin{figure}
\centering
\includegraphics[width=0.4 \textwidth]{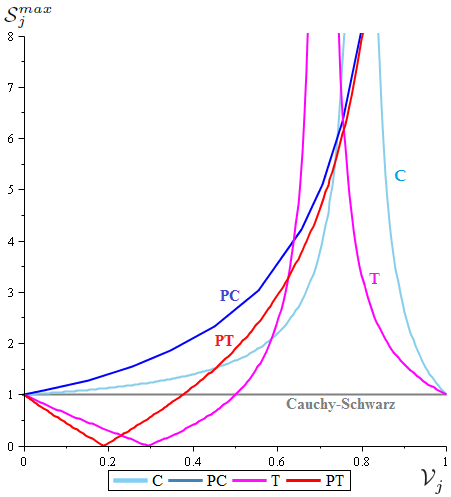}
\caption{\label{csplot} Plot of the maximum of the Schwarz function ${\cal S}^{max}_{{ j}}$ dependent on the visibility ${\cal V}_j$ ($j = \{T,PT,C,PC\}$) for combinations of quantum and classical photon sources (cf.~Eq.~(\ref{schwarzeq}) and text).}
\end{figure}


The corresponding behavior of the Schwarz function ${\cal S}^{max}_{{ j}}$ as a function of the visibilities ${\cal V}_{{j}}$ ($j = \{T,PT,C,PC\}$) is displayed in Fig.~\ref{csplot}. With the pure classical signal it is not possible to violate the CS inequality as ${\cal S}^{max}_{{ Class}} = 1$ for all ${\cal V}_{{Class}}$. For a mixed quantum-thermal signal (T) the CS inequality is only violated for a visibility ${\cal V}_T > 50\%$, i.e., if the net photon number in the thermal beam fulfills $\bar{n} < 1$ (cf.~Eq.~\ref{VQMT}). Note that a visibility $> 50\%$ is just above the maximal visibility of the pure classical coherent signal (cf.~Eq.~(\ref{VCC})). Moreover, for ${\cal V}_T = 1/\sqrt{2}$ the CS function ${\cal S}^{max}_{{ T}}$ diverges. By contrast, if we consider the mixed quantum-PATS signal (PT), the CS inequality can be violated already for a visibility ${\cal V}_{{ PT}} > 37.5\%$, corresponding to a net photon number $\mathfrak{n}_{PT} < 3$ of source $B$ (cf.~Eq.~(\ref{VQMPATSnet})). Furthermore, we have ${\cal S}^{max}_{{ PT}} \rightarrow \infty$ for ${\cal V}_{{ PT}} \rightarrow 1$. 

Considering the mixed quantum-coherent signal (C) and the mixed quantum-PACS signal (PC), if we choose a phase difference of $\pi$ between $\varphi_{1}$ and $\varphi_{2}$, we have a violation of the CS inequality for any finite net photon number of source $B$, what corresponds to a required visibility of these signals $> 0$ (see~Fig.~\ref{visQMC} and Eqs.~(\ref{VQMC}), (\ref{VQMPACS}) and (\ref{nPACS})). For the mixed quantum-coherent signal (C) we find ${\cal S}^{max}_{{ C}} \rightarrow \infty$ for a visibility ${\cal V}_{{ C}} = 80\%$, corresponding to a net photon number of $\bar{n} = 1/2$ in the coherent beam (cf.~Eq.~(\ref{VQMC})). For the mixed quantum-PACS signal (PC) the CS function goes to $\infty$ for ${\cal V}_{{ PT}} \rightarrow 1$ (see Fig.~\ref{visQMC}). In this way we can pin down quantitatively the transgression to the non-classical regime for all our investigated mixed quantum-classical and mixed quantum-semiquantum signals. As expected, a signal of two single photon sources also violates the CS inequality, what is not depicted in Fig.~\ref{csplot}. Finally it should be borne in mind that from no violation of the CS inequality no conclusion can be drawn regarding the quantum nature of the radiation field.


\section{Test of the non-local character of the mixed radiation fields}
\label{schsh}




In the past there have been many tests of Bell inequalities using photons obtained from parametric down converters (e.g., \cite{Mandel87,Shih88,Zeilinger93,Zeilinger95,Teich07}). Further Bell inequalities have been studied using photons obtained either from two independent down converters \cite{Zeilinger93} or one photon from a downconverter and one photon from a coherent source \cite{QMxC1}. In all these experiments Bell inequalities were tested using polarization (spin like) correlations between the photons. We have already seen in the previous section that the radiation field produced by two disparate sources, where one source is a single photon emitter, is strictly quantum in nature in certain ranges of the net photon number of source $B$. Hence we would expect a violation of Bell-type inequalities \cite{wiegner}. However, in order to see this in detail we need to normalize our photon-photon correlation funcion. We use the following normalization (see~Appendix \ref{calc})

\staf
\tilde{g}^{(2)}(\varphi_1, \varphi_2) = \frac{G^{(2)}(\varphi_1, \varphi_2)}{{\cal N}} - 1\,,
\label{norm}
\stof
\noindent where ${\cal N}$ is an appropriate scaling factor. This scaling factor can be obtained by physical considerations - we take it to be the integrated photon-photon correlation function obtained by fixing one detector and moving the other detector. Using the moments given in Eqs.~(\ref{momc}), (\ref{momt}), (\ref{momspacs}) and (\ref{momspats}), and recalling that we assume uncorrelated sources and that we have $\left\langle I_A^2 \right\rangle_{QM} = 0$, we can express the normalized photon-photon correlation functions in the collective form 


\staffeld
\label{G2main}
\tilde{g}^{(2)}_{j}(\varphi_1, \varphi_2) = \frac{G^{(2)}_{{j}}(\varphi_1, \varphi_2)}{2\,\pi^{-1}\int G^{(2)}_{{j}}(\varphi_1, \varphi_2) d\varphi_1 |_{\varphi_2 = const.}} - 1 = \nonumber\\
= \frac{\left\langle (I_A +I_B)^2\right\rangle +2\,\left\langle I_A\right\rangle \left\langle I_B\right\rangle\cos{(\varphi_2 - \varphi_1)}}{\left\langle (I_A + I_B)^2\right\rangle} -1 = \nonumber\\
= {\cal V}_{{j}} \, \cos{(\varphi_2 - \varphi_1)}\,,\hspace{4.65cm}
\stoffeld

\noindent with the visibilities ${\cal V}_{{j}}$ given in Eqs.~(\ref{VCC}), (\ref{VQMC}), (\ref{VQMT}), (\ref{VQMPACS}) and (\ref{VQMPATSnet}), respectively. We want to emphasize that our proposed rescaled correlation functions are independent of experimental inefficiencies, e.g., the quantum efficiency of the detectors and the restricted angle subtended by the detector surfaces (see~Appendix \ref{calc}). Eq.~(\ref{G2main}) is the celebrated correlation commonly used in testing violations of Bell-type inequalities \cite{Bell72}. This is also the correlation for two spins in a Werner state which is a mixed entangled state \cite{Werner89}. It is well known that such a correlation violates Bell's inequalities in the CHSH formulation \cite{chsh}

\staffeld
|\tilde{g}^{(2)}_{j}(\varphi_1, \varphi_2) - \tilde{g}^{(2)}_{j}(\varphi_1, \varphi_2') + \hspace{1.5cm}\nonumber\\
\hspace{1.5cm}+\, \tilde{g}^{(2)}_{j}(\varphi_1', \varphi_2) + \tilde{g}^{(2)}_{j}(\varphi_1', \varphi_2')| \leq 2
\stoffeld

\noindent if 

\staf
{\cal V}_{{j}} > \frac{1}{\sqrt{2}}\;.
\stof

\noindent Down to the present day this particular version of Bell's inequalities is tested in most experiments  (cf., e.g., \cite{QMxC1,Shih88,Zeilinger93,Zeilinger95}). From Fig.~\ref{visQMC} and Eqs.~(\ref{VQMC}), (\ref{VQMT}), (\ref{VQMPACS}), (\ref{nPACS}) and (\ref{VQMPATSnet}), we can derive the net photon numbers $\mathfrak{n}_j$ of source $B$ for which the CHSH inequalities are violated: The net photon number must be less than (a) $\sqrt2 - 1$ for a thermal source, (b) $\approx 1.45$ for a photon added thermal source ($\bar{n} \approx 0.22$), (c) $2\,(\sqrt2 - 1)$ for a coherent source and (d) $\approx 1.52$ for a photon added coherent source ($\bar{n} \approx 0.29$). Clearly in order to see the violations of locality the condition on $\mathfrak{n}_j$ for photon added sources is more relaxed.

\section{Summary}
\label{summary}

In conclusion we studied position dependent photon-photon correlations of photons emitted from disparate mixed quantum-classical and mixed quantum-semiquantum sources in free space. As quantum source we considered a spontaneously emitting atom; however we want to emphasize that our results are not limited to this case as any single photon source can act as a quantum source in the presented scheme \footnote{The experiments of refs \cite{QMxC1,QMxC2,QMxT} use a single photon obtained from a down converter.}. The classical sources were represented either by thermal or coherent beams. These two classical sources have been studied in the past by overlapping photons at a beam splitter \cite{QMxC1,QMxT}. As semiquantum source we investigated both, PATS and PACS, what corresponds to a whole new class of disparate mixed quantum photon sources at the quantum-classical boundary. We introduced the CS inequality and obtained quantitative results for the non-classicality of all our mixed radiation fields, depending on the visibility ${\cal V}_j$ of the corresponding photon-photon correlations (see Fig.~\ref{csplot}), what relates also to the net photon number $\mathfrak{n}_{j}$ in the field of source $B$ (see Fig.~\ref{visQMC}). By normalizing the photon-photon correlation function in a physical way we obtained a violation of Bell-type inequalities for all considered mixed fields, again dependent on the visibility ${\cal V}_j$ and the net photon number $\mathfrak{n}_{j}$ of source $B$. The violation of Bell-type inequalities requires more stringent conditions on the visibility of the photon-photon correlations than the violation of the CS inequality. This is reminiscent of the behavior of Werner states for spins.

\section{Ackowledgements}

R.W.~gratefully acknowledges financial support by the Elite Network of Bavaria. This work was supported by the Deutsche Forschungsgemeinschaft. 

\appendix

\section{CS inequality for photon-photon correlations}
\label{csapp}

Let us investigate Eq.~(\ref{trrhodeltai}) in the form

\staf
\label{trrhodeltai1}
Tr\left\{ \rho\, [ \alpha \,\delta \hat{I}(\varphi_1) + \beta \,\delta \hat{I}(\varphi_2) ]^2 \right\} \geq 0 \;\, \forall\, \alpha,\beta\;,
\stof

\noindent with $\delta \hat{I}(\varphi_i) \equiv \hat{I}(\varphi_i) - \langle \hat{I}(\varphi_i)\rangle$ and the density operator $\rho$ in the Glauber-Sudarshan P-representation

\staf
\rho = \int P(\alpha) \ket{\alpha}\bra{\alpha} d Re(\alpha)d Im(\alpha)\;.
\stof

\noindent $P(\alpha)$ abbreviates a (quasi-)probability function depending on the state of the investigated light field. For classical states of light $P(\alpha)$ is a classical probability distribution assuming only positive values. Since for any  $\alpha$ and $\beta$ the quantity $[ \alpha \delta \hat{I}(\varphi_1) + \beta \delta \hat{I}(\varphi_2) ]^2$ is also positive, inequality (\ref{trrhodeltai1}) holds for any classical light field. However, for nonclassical states $P(\alpha)$ is a quasi-probability distribution and the P-function can be negative, singular or need not exist. Thus, the inequality can be violated.

From Eq.~(\ref{trrhodeltai1}) it follows

\staf
\begin{vmatrix} \langle \delta \hat{I}(\varphi_1)\,\delta \hat{I}(\varphi_1) \rangle & \langle \delta \hat{I}(\varphi_1)\,\delta \hat{I}(\varphi_2) \rangle \\ \langle \delta \hat{I}(\varphi_2)\,\delta \hat{I}(\varphi_1) \rangle & \langle \delta \hat{I}(\varphi_2)\,\delta \hat{I}(\varphi_2) \rangle \end{vmatrix} \geq 0\;.
\stof

\noindent Evaluating the determinant we arrive at

\staffeld
\prod_{i = 1}^2 \langle  \hat{I}(\varphi_i)\hat{I}(\varphi_i) - 2\,\hat{I}(\varphi_i)\,\langle \hat{I}(\varphi_i)\rangle + \langle \hat{I}(\varphi_i)\rangle\langle \hat{I}(\varphi_i)\rangle\rangle - \nonumber\\
- \prod_{i,j = 1, i \neq j}^2 \langle \hat{I}(\varphi_i)\hat{I}(\varphi_j) - \hat{I}(\varphi_i)\,\langle \hat{I}(\varphi_j)\rangle -\hspace{1.5cm}\nonumber\\
- \;\hat{I}(\varphi_j)\,\langle \hat{I}(\varphi_i)\rangle + \langle \hat{I}(\varphi_i)\rangle \langle \hat{I}(\varphi_j)\rangle \rangle\;.\hspace{0.5cm}
\stoffeld

\noindent If we now use the identities

\staffeld
\langle \hat{I}(\varphi_i)\rangle &=& \langle I_A + I_B \rangle\  \,,\nonumber\\
\langle \hat{I}(\varphi_i)\hat{I}(\varphi_j) \rangle &=&  G^{(2)}(\varphi_i,\varphi_j) \,,
\stoffeld

\noindent we obtain Eq.~(\ref{schwarzeq}).\\



We want to emphasize that if instead of Eq.~(\ref{trrhodeltai1}) we use arguments similar to those leading to Eq.~(\ref{27}) we would get a CS inequality for normalized photon-photon correlation functions $g^{(2)}(\varphi_i, \varphi_j)$ of the form (cf., e.g., \cite{walls86})

\staf
|g^{(2)}(\varphi_1, \varphi_2)|^2 \leq g^{(2)}(\varphi_1,\varphi_1)\,g^{(2)}(\varphi_2,\varphi_2)\,.
\stof

\noindent For two quantum sources this form of the CS inequality is never violated as

\staf
g^{(2)}(\varphi_1, \varphi_2) = \frac{1}{2}\,(1 + \cos{(\varphi_2 - \varphi_1)})\,.
\stof 

Thus, clearly such an inequality does not enable us to test the quantum nature of our sources. 



\section{Derivation of position dependent Bell correlations}
\label{calc}

In this appendix we want to derive Eqs.~(\ref{norm}) and (\ref{G2main}). Bell inequalities in the CHSH formulation are inequalities based on a \emph{double channel} experiment. In order to establish this double channel character, we have to slightly change our setup. Let us consider that the photon sources $A$ and $B$ are connected to both detectors via optical fibers. Each \emph{detector} shall now consist of two photomultipliers, located at the output ports of a 50-50 beamsplitter, representing a transmitting $+$ and a reflecting $-$ channel.  Measuring the photon-photon correlation function, we thus consider four photodetectors, say $D_1^+,D_1^-,D_2^+$ and $D_2^-$. In this configuration we can make use of the well known correlation function \cite{Bell72,yurke88} commonly employed in Bell experiments

\staf
{\cal C}(\varphi_1,\varphi_2) \equiv \frac{\left\langle \sum_{\alpha,\beta} \alpha\,\beta\,G^{(2)}_{{\alpha\,\beta}}(\varphi_1, \varphi_2) \right\rangle}{\left\langle \sum_{\alpha,\beta}\,G^{(2)}_{{\alpha\,\beta}}(\varphi_1, \varphi_2)  \right\rangle}\,,
\stof

\noindent where $\alpha,\beta = \left\{ +,-\right\}$. Hereby, $G^{(2)}_{{\alpha\,\beta}}(\varphi_1, \varphi_2)$ abbreviates the photon-photon correlation of measuring one photon in the $\alpha$ channel at position $\varphi_1$ (i.e., at detector $D_1^\alpha$) and the other photon in the $\beta$ channel at position $\varphi_2$ (i.e., at detector $D_2^\beta$). In an experiment these correlations would correspond to four twofold coincidence rates. Keeping in mind that due to reflection we get a phase shift of $\pi$ we arrive at

\staf
G^{(2)}_{{\alpha\,\beta}}(\varphi_1, \varphi_2) = \left\langle (I_A +I_B)^2\right\rangle \pm 2\,\left\langle I_A\right\rangle \left\langle I_B\right\rangle\cos{(\varphi_2 - \varphi_1)}\,,
\stof

\noindent where the plus (minus) sign holds for $\alpha = \beta$ ($\alpha \neq \beta$). Thus, ${\cal C}(\varphi_1,\varphi_2)$ calculates to

\staf
\label{appendix2}
{\cal C}(\varphi_1,\varphi_2) = {\cal V}_{{j}} \, \cos{(\varphi_2 - \varphi_1)}\,,
\stof

\noindent with the visibility 

\staf
{\cal V}_{{j}} = \frac{2\,\left\langle I_A \right\rangle \left\langle I_B\right\rangle}{\left\langle (I_A + I_B)^2\right\rangle}\,.
\stof

\noindent Eq.~(\ref{appendix2}) is equivalent to Eq.~(\ref{G2main}), i.e., Eq.~(\ref{G2main}) integrates the foregoing procedure of deriving correlations suitable for a Bell test. We want to emphasize that this expression holds for \emph{every} position dependent mixed photon-photon correlation function with visibility ${\cal V}_{{j}}$. Due to the normalization on $\left\langle \sum_{\alpha,\beta}\,G^{(2)}_{{\alpha\,\beta}}(\varphi_1, \varphi_2) \right\rangle$ the correlation ${\cal C}(\varphi_1,\varphi_2)$, like Eq.~(\ref{norm}), becomes independent of experimental inefficiencies of the form \mbox{$\eta = \alpha_0 \frac{\Delta \Omega}{4\pi}$} which we have set to unity in the course of this paper. Hereby, $\alpha_0$ stands for the quantum efficiency of the detectors and $\Delta \Omega$ for the restricted solid angle subtended by the detector surfaces. 






\end{document}